\documentclass[sigconf,10pt,nonacm]{acmart}

\usepackage{balance}

\usepackage[utf8]{inputenc}%

\usepackage{amsfonts}

 \usepackage{amsmath}
\usepackage{comment}
\usepackage{cancel}

\usepackage{ifthen}

\usepackage{tikz}

\def\BibTeX{{\rm B\kern-.05em{\sc i\kern-.025em b}\kern-.08emT\kern-.1667em\lower.7ex\hbox{E}\kern-.125emX}}

\copyrightyear{2019} 
\acmYear{2019} 
\setcopyright{acmcopyright}
\acmConference[PODS'19]{38th ACM SIGMOD-SIGACT-SIGAI Symposium on Principles of Database Systems}{June 30-July 5, 2019}{Amsterdam, Netherlands}
\acmBooktitle{38th ACM SIGMOD-SIGACT-SIGAI Symposium on Principles of Database Systems (PODS'19), June 30-July 5, 2019, Amsterdam, Netherlands}
\acmPrice{15.00}
\acmDOI{10.1145/3294052.3319693}
\acmISBN{978-1-4503-6227-6/19/06}

\begin{document}

\title{Compiling Existential-Positive Queries to Bounded-Variable Fragments}

\author{Christoph Berkholz}
\authornote{Funded by the Deutsche Forschungsgemeinschaft (DFG, German Research
Foundation) -- SCHW~837/5-1.}
\affiliation{Humboldt-Universität zu Berlin \\
  Berlin, Germany\\
}
\email{berkholz@informatik.hu-berlin.de}

\author{Hubie Chen}
\authornote{Acknowledges the support of Spanish Project TIN2017-86727-C2-2-R.}
\affiliation{Dept.\ of Informatics\\
King's College London\\
London, United Kingdom\\
}
\email{hubie@kcl.ac.uk}

\renewcommand{\shortauthors}{Berkholz and Chen}

\begin{abstract}
A crucial property of bounded-variable fragments of first-order logic
is that they can be evaluated in polynomial time. It is therefore a
useful preprocessing step to rewrite, if possible, a first-order query
to a logically equivalent one with a minimum number of
variables. However, it may occur that reducing the number of
variables causes an increase in formula size.
We investigate this trade-off for the existential-positive fragment of
first-order queries, where variable minimisation is decidable in
general.  In particular, we study the blow-up in the formula size when
compiling existential-positive queries to the bounded variable
fragment of positive first-order logic.  While the increase of the
formula size is always at most exponential, we identify situations
(based on the signature and the number of variables) where only a
polynomial blow-up is needed.  In all other cases, we show that an
exponential lower bound on the formula size of the compiled formula
that matches the general upper bound.
This exponential lower bound is unconditional, and is the first
unconditional lower bound for formula size with respect to the studied
compilation; it is proved via establishing a novel interface with 
circuit complexity which
may be of future interest.
\end{abstract}




\maketitle

\newcounter{CommentCounterRed}
\newcounter{CommentCounterBlue}
\newcounter{CommentCounterGreen}
\newcounter{CommentCounterGray}

\newcommand{\writecommentRed}[2]{\textcolor{red}{\stepcounter{CommentCounterRed} \textbf{#1 \arabic{CommentCounterRed}:} #2}}
\newcommand{\writecommentBlue}[2]{\textcolor{blue}{\stepcounter{CommentCounterBlue} \textbf{#1 \arabic{CommentCounterBlue}:} #2}}
\newcommand{\writecommentGreen}[2]{\textcolor{green}{\stepcounter{CommentCounterGreen} \textbf{#1 \arabic{CommentCounterGreen}:} #2}}
\newcommand{\writecommentGray}[2]{\textcolor{gray}{\stepcounter{CommentCounterGray} \textbf{#1 \arabic{CommentCounterGray}:} #2}}

\newtheorem{prop}[theorem]{Proposition}
\newtheorem{observation}[theorem]{Observation}
\newtheorem{remark}[theorem]{Remark}
\newtheorem{examplecore}[theorem]{Example}

\newcommand{\setc}[2]{\{ #1 ~|~ #2 \}}
\newcommand{\graphG}{\mathbf{G}}
\newcommand{\AG}{\relA_{\graphG}}
\newcommand{\tuple}[1]{\ensuremath{\mathbf{#1}}}
\newcommand{\tupv}{\tuple{v}}
\newcommand{\subst}[3]{\ensuremath{#1\tfrac{#2}{#3}}}
\newcommand{\formsizeN}{\ensuremath{N}} %
\newcommand{\sigunary}{{\sigma^m_U}}

\renewcommand{\epsilon}{\varepsilon}
\renewcommand{\phi}{\varphi}

\newcommand{\rela}{\mathbf{A}}
\newcommand{\relb}{\mathbf{B}}
\newcommand{\relc}{\mathbf{C}}
\newcommand{\reld}{\mathbf{D}}

\newcommand{\fancya}{\mathcal{A}}
\newcommand{\fancyc}{\mathcal{C}}
\newcommand{\fancyl}{\mathcal{L}}
\newcommand{\fancym}{\mathcal{M}}
\newcommand{\fancyr}{\mathcal{R}}

\newcommand{\restr}{\upharpoonright}

\renewcommand{\matrix}{\mathrm{matrix}}
\newcommand{\ar}{\mathrm{ar}}

\renewcommand{\S}{\mathcal{S}}

\newcommand{\entail}{\mathsf{ENTAIL}}
\newcommand{\finentail}{\mathsf{FIN}\textup{-}\mathsf{ENTAIL}}

\newcommand{\disj}{\mathrm{Disj}}

\newcommand{\image}{\mathsf{image}}

\newcommand{\numforall}{\sharp \forall}
\newcommand{\numexists}{\sharp \exists}

\newcommand{\numu}{\sharp u}
\newcommand{\nume}{\sharp e}

\newcommand{\free}{\mathsf{free}}
\newcommand{\appears}{\mathsf{appears}}
\newcommand{\size}{\mathsf{size}}

\newcommand{\EP}[1]{\ensuremath{\mathrm{EP}^{#1}}}
\newcommand{\exEP}[1]{\ensuremath{\mathord{\approx}\mathrm{EP}^{#1}}}

\newcommand{\PP}[1]{\ensuremath{\mathrm{PP}^{#1}}}
\newcommand{\exPP}[1]{\ensuremath{\mathord{\approx}\mathrm{PP}^{#1}}}

\newcommand{\PFO}[1]{\ensuremath{\mathrm{PFO}^{#1}}}

\newcommand{\FO}[1]{\ensuremath{\mathrm{FO}^{#1}}}
\newcommand{\exFO}[1]{\ensuremath{\mathord{\approx}\mathrm{FO}^{#1}}}
\newcommand{\NP}{\ensuremath{\mathrm{NP}}}

\newcommand{\real}{\ensuremath{\operatorname{real}}}

\newcommand{\N}{\mathbb{N}}
\newcommand{\nats}{\N}

\newcommand{\fpt}{\mathsf{FPT}}
\newcommand{\ptime}{\mathsf{PTIME}}
\newcommand{\fptime}{\mathsf{fPTIME}}
\newcommand{\np}{\mathsf{NP}}
\newcommand{\conp}{\mathsf{co\textup{-}NP}}
\newcommand{\nc}{\mathsf{NC}}

\newcommand{\chopped}[1]{\mathsf{chopped\textup{-}}#1}
\newcommand{\mixed}[1]{\mathsf{mixed\textup{-}}#1}

\newcommand{\mathcomp}{\mathsf{\textup{-}comp\textup{-}}}
\newcommand{\lcompc}{\fancyl\mathcomp\fancyc}
\newcommand{\lcompfptime}{\fancyl\mathcomp\fptime}
\newcommand{\polycomp}[1]{\mathsf{poly\textup{-}comp\textup{-}}#1}
\newcommand{\expcomp}[1]{\mathsf{exp\textup{-}comp\textup{-}}#1}
\newcommand{\polycompp}{\polycomp{\ptime}}

\newcommand{\qe}{\mathsf{QueryEval}}

\newcommand{\wpfin}{\wp_{\mathrm{fin}}}

\newcommand{\clique}{\mathsf{CLIQUE}}
\newcommand{\nuclique}{\mathsf{\nu\textup{-}\clique}}

\section{Introduction}
\label{sect:introduction}

\newcommand{\EF}{Ehrenfeucht-Fra\"iss\'e}

{\bf Overview.}
In this work we study the size blow-up that might occur when
rewriting an existential-positive first-order query into a logically equivalent
one that has a bounded number of variables.
Existential-positive queries (\EP{}) are those first-order queries that only
use existential quantification ($\exists$),
conjunction  ($\land$),
and disjunction ($\lor$)--so, those that disallow negation and
universal quantification.
This query language has been studied intensively in database theory
and constraint satisfaction~\cite{Chen-tocl,ChenMengel16-pods-ep-counting,ChenMengel17-lics-logic-counting}.
It can be
alternatively described by the fragment of relational algebra that uses Select,
Project, Join, and Union (SPJU), which is in turn semantically equivalent to the
class of unions of conjunctive queries (UCQ).

{\bf Measures and their motivations.}
In addition to the \emph{size} of a query, a fundamental complexity measure of first-order queries is the \emph{number of variables}, 
by which is meant the total number of different variables that occur throughout
a particular first-order query.
The number of variables of query is essentially equivalent to the 
\emph{width} of a query, which is defined to be the maximum number of free variables
over all subformulas of the query.  A query using $k$ variables clearly has
width at most $k$, and in the other direction, a query having width $k$
can be rewritten, just by the syntactic renaming of variables,
to a query using $k$ variables.  We thus freely interchange between these
two measures in our discussions.

Every first-order query $\varphi$ of width $k$ can be evaluated on a database of
size $n$ in time $O(\|\varphi\|\cdot k\cdot n^k)$ by a
straightforward bottom-up algorithm \cite{Vardi95} that materialises
for each subformula the result relation, which is of arity at most $k$.
Therefore, it is a useful preprocessing step in query optimisation to
rewrite, if possible, a first-order query into one having smaller,
or even minimal, width.
A classical example of a query where the width can be decreased
via rewriting is the path query
$$
\varphi(x,y) = 
\exists z_1\cdots
\exists z_k
\bigl(
E(x,z_1)\land
E(z_1,z_2)\land
\cdots 
$$
$$
\land
E(z_{k-1},z_k) \land
E(z_{k},y)
\bigr),$$ which can be equivalently formulated as the formula
$$
\psi(x,y) = 
\exists z
\Bigl(
E(x,z)\land
\exists x
\bigl(
E(z,x)\land
\exists z
(
E(x,z)\land
\cdots
\land
E(z,y)
)
\bigr)
\Bigr)
$$
which uses $3$ variables, and is obtained from $\varphi(x,y)$
by pushing in the quantifiers and renaming bounded variables.
It is known 
  that one can always compute, from
an existential-positive query, 
an equivalent existential-positive query of minimal
width~\cite{BovaChen14,BovaChen15-eplogic}. 
In fact, the existential-positive queries form a
maximal fragment where this is computable: it is folklore that for full
first-order logic width minimisation is undecidable, and the same holds
true for positive first-order logic ($\PFO{}$), 
which is the fragment of first-order logic where 
negation is disallowed~\cite{BovaChen14}.

From the viewpoint of parametrized complexity, width minimisation
with a subsequent bottom-up evaluation is essentially \emph{the only} way to
achieve tractable query evaluation for existential-positive queries:
model-checking for classes of existential-positive queries is
fixed-parameter tractable\footnote{Parameterized by the query size and
  assuming that the arity of each relation is bounded above by a constant.} 
if and only
every query in that class is equivalent to one of bounded width \cite{Chen-tocl}. 

All together, this provides strong motivation for the study of 
rewriting existential-positive queries to
logically equivalent ones of bounded width.
Such a rewriting can always be done 
as follows~\cite{SagivYannakakis80-equivalences,BovaChen14}: First transform the
existential-positive formula into prenex normal form. Afterwards
translate the quantifier-free part into a positive DNF (a disjunction of
conjunctions of non-negated atoms) and push the
existential quantifiers over the disjunction to obtain a disjunction of
conjunctive queries, which is a UCQ. By minimising this UCQ using
standard techniques, one can obtain an existential-positive formula of
minimal width.
This approach has one crucial drawback: due to the transformation into
DNF form, the size of the query might increase exponentially!
The main issue addressed in this paper is whether, and when, this exponential
increase of the query size is necessary.

\subsection{Our contribution}

Based on the signature\footnote{In this paper, a \emph{signature}
  (sometimes called \emph{schema}) will
always be a (not necessarily finite) set of relation symbols. For
completeness, we also allow nullary relation symbols, which are
interpreted as Boolean variables.} and the minimum width $k$ of an
equivalent query, we identify all cases where the exponential blow-up
for width minimisation
can be avoided.  In these situations, it is always possible
to compile an existential-positive formula into one that has minimum
width and where the size is only polynomial %
in the size of the original query.

In all other cases, we show that there are families of
existential-positive formulas $\varphi_n$ of size $n$, which are
equivalent to existential-positive $k$-variable formulas, but where
any such $k$-variables formula have size $\Omega\bigl(2^{n^\varepsilon}\bigr)$ for
some absolute constant $\epsilon > 0$.  Our exponential lower bounds
even hold when compiling formulas into equivalent positive formulas,
which can also make use of universal quantification ($\forall$) and can have
any finite number of variables.  The upper and the matching lower
bounds lead to a full characterisation of existential-positive queries
where polynomial size compilation to negation-free bounded variable fragments is
possible. We formalise the characterisation theorems in
Section~\ref{sec:classificationtheorems}.
The core of our lower bound argument builds on a rather special
encoding of the clique problem in existential-positive logic. 
Our formula size lower bounds then follow from a
translation from bounded-variable formulas to circuits, in conjunction with
a novel use of the well-known lower bound for monotone Boolean
circuits \cite{Razborov85}.

So, we show unconditional lower bounds on the size blow-up incurred
by compilations from sentences to bounded-variable fragments,
where a compilation is simply a mapping from sentences to sentences
which preserves logical equivalence.
We also complement these unconditional lower bounds with
conditional lower bounds on the size blow-up incurred
by compilations from sentences to arbitrary formats
which then aid a polynomial-time decision procedure;
these conditional lower bounds are proved alongside the unconditional lower bounds,
and a summary and discussion thereof is provided in 
Section~\ref{sect:param-comp-summary}.

\subsection{Related work}

Closely related to our main goal of understanding the compilation to
bounded variable fragments of existential-positive logic is the work
of 
Bova and the second author of this paper~\cite{BovaChen14}, who investigated the
computational complexity of deciding whether a given \EP{} formula can
be transformed into an equivalent one of  width $k$.
They showed that
this decision problem is complete for the second level of the
polynomial hierarchy. It follows that, unless $\ptime = \np$, there is no
polynomial time algorithm that rewrites an \EP{} query into one of width $k$.
Moreover, under the reasonable assumption
that the polynomial hierarchy does not collapse, it is not possible to
compile an \EP{} query into a bounded-width query of size polynomial in the
original query.
Note that while this result implies a super-polynomial lower bound
based on complexity theoretic assumptions, in this paper we are able
to prove an explicit $\Omega\bigl(2^{n^{1/14}}\bigr)$ lower bound on formula
size that does not rely on any assumptions.

Also related to the main theme of our work is the notion of \emph{succinctness}
in finite model theory, which measures how much more succinctly
properties can be expressed in one logic compared to another logic.
Two logics $\fancyl$ and $\fancyl'$ are usually
compared by their expressiveness: $\fancyl$ is at least as expressive as
$\fancyl'$ if every property that can be expressed in $\fancyl'$ can
also be expressed in $\fancyl$.
Succinctness additionally takes the size of a formula into account:
$\fancyl$ is at least as succinct as $\fancyl'$ if every property that
can be expressed in $\fancyl'$ can
also be expressed in $\fancyl$ by a sentence of roughly the same
size. Moreover, $\fancyl$ is exponentially more succinct as $\fancyl'$
if in addition there is some $\varphi\in\fancyl$ that expresses a property
expressible in $\fancyl'$, but for which the size of every $\psi\in \fancyl'$ that
is equivalent to $\varphi$ is at least exponential in the size of $\varphi$.

Contrary to expressiveness and the well-studied complexity measures
width, quantifier depth, and number of quantifier alternations, not
that much is known about succinctness and the size of formulas.  Wilke
\cite{Wilke99} showed that CTL$^+$ is exponentially more succinct that
CTL and subsequent improvements of this succinctness gap have been
obtained by Adler and Immerman \cite{AdlerImmerman03} as well as Lange \cite{Lange08}.  In particular,
Adler and Immerman \cite{AdlerImmerman03} developed a combinatorial
method based on \EF{} games for proving formula size lower bounds and
applied it to prove an optimal succinctness gap between CTL$^+$ and
CTL. Grohe and Schweikardt \cite{GroheSchweikardt05} used a related
method to prove that first-order logic is exponentially more succinct
than its 3-variable fragment FO$^{3}$ by showing that a linear order
of length $n$ can be defined by a 4-variable first-order formula of
size $O(\log n)$, but requires FO$^3$ formulas of size
$\Omega(\sqrt{n})$.
Further succinctness results have been obtained for 2-variable logic
on words \cite{WeisPhd2011} and for first-order and monadic
second-order logic on structures of bounded tree-depth \cite{EickmeyerElberfeldHarwath17}.
Here, we can also mention a qualitatively related succinctness result on 
navigational queries for graphs~\cite[Theorem 6.3]{Surinx-navigate}.

Stated in terms of succinctness, our results imply that, for each
$k\geq 1$, existential-positive first-order logic (\EP{}) as well as
positive first-order logic (\PFO{}) are exponentially more succinct
than their respective $k$-variable fragments \EP{k} and \PFO{k}.
For $k\geq 4$ it is open, whether full first-order logic (\FO{}) and is
exponentially more succinct than its  $k$-variable fragment $\FO{k}$.

The technical core of our compilation lower bound relies on a novel
use of the well-known connection between descriptive complexity and
Boolean circuit complexity.
It is folklore that the circuit class DLOGTIME-uniform AC$^0$
corresponds to FO[$\leq,+,\times$],
i.\,e., a property of relational structures can be computed by a
uniform family of constant depth circuits, if and only if, it can be
expressed in first-order logic with a linear order and the arithmetic
predicates $+$ and $\times$ on the universe.

From this connection it follows that 
circuit lower bounds for AC$^0$ imply inexpressibility results for
first-order logic, even in the presence of arithmetic predicates.
Moreover, a fixed $k$-variable first-order formula can be translated to a family of
AC$^0$ circuits of size $O(n^k)$ that decides the same problem.
Hence, more fine-grained lower bounds for AC$^0$ provide
inexpressibility results for the $k$-variable fragment.
In his ground-breaking work \cite{Rossman08}, Rossman showed an $\omega(n^{k/4})$ AC$^0$ lower bound for
$k$-clique and used the connection to descriptive complexity in order to show that FO$^{k+1}$ has a strictly stronger
expressiveness than FO$^{k}$ on ordered graphs.
Note that while from \cite{GroheSchweikardt05} it follows that FO$^4$ is
exponentially more succinct than FO$^{3}$ on ordered structures, for
$k\geq 4$ it is open whether there is a succinctness gap between
FO$^k$ and FO$^{k-1}$ (even on the class of all structures).

We establish our succinctness results for $\EP{}$ and $\PFO{}$ through
circuit lower bounds for \emph{monotone} Boolean circuits and it is
crucial for us that, contrary to the general case, there are lower
bounds for monotone circuits of arbitrary depth \cite{Razborov85}. 
While these size lower bounds for monotone circuits have been fruitfully
applied in related ares, e.\,g., for
query rewriting in ontology-based data access
\cite{GottlobKKPSZ14}, we believe that its application to variable minimisation is a  technical contribution might be of
independent interest.

\section{Preliminaries}

\subsection{Logic and structures}

We formalize our results using relational first-order logic,
which we now review briefly.

A signature $\sigma$ is a set of relation symbols, each of which has an associated arity;
we permit each natural number $k \geq 0$ as an arity.
We refer to a relation symbol of arity $0$ as a \emph{propositional symbol},
and a relation symbol of arity $1$ as a \emph{unary symbol}.
We refer to a relation symbol that is of arity $2$ or greater as
an \emph{at least binary symbol}.

A \emph{structure} $\relb$ over a signature $\sigma$ consists of a non-empty set $B$
called the \emph{universe} of the structure,
and, for each $R \in \sigma$, a relation $R^{\relb}$ over $B$ whose arity is that of $R$.
We generally use the letters $\rela, \relb, \ldots$ to denote structures,
and the corresponding letters $A, B, \ldots$ to denote their respective universes.
When $\rela$ and $\relb$ are structures over the same signature $\sigma$,
we define their union, denoted by $\rela \cup \relb$,
as the structure with universe $A \cup B$
and where, for each relation symbol $R \in \sigma$,
it holds that $R^{\rela \cup \relb} = R^{\rela} \cup R^{\relb}$.

An \emph{atom} (over signature $\sigma$) is an equality of variables
$u=v$ or a predicate application $R(v_1, \ldots, v_r)$ where $R \in \sigma$
is a relation symbol of arity $r$, and the $v_i$ are variables.
A predicate application $P()$ consisting of a relation symbol $P$ of arity $0$
is referred to as a \emph{propositional atom}.
A \emph{formula} (over signature $\sigma$) is built from atoms (over $\sigma$),
negation ($\neg$), conjunction ($\wedge$), disjunction ($\vee$), 
universal quantification ($\forall$), and existential quantification ($\exists$).
We will sometimes use $\top$ and $\bot$ to denote the values
true and false, respectively.
We use $\free(\phi)$ to denote the set of free variables of a formula $\phi$;
a \emph{sentence} is a formula having no free variables.

When $\phi$ is a formula and the notation $\phi(x_1, \ldots, x_k)$ is used, 
it is assumed that the free variables of $\phi$ are contained in $\{ x_1, \ldots, x_k \}$.
When $\phi(x_1, \ldots, x_k)$ is a formula
and $\relb$ is a structure,
we use the notation $\phi(\relb)$ to denote the set of all tuples
$(b_1, \ldots, b_k) \in B^k$ such that
$\relb, b_1, \ldots, b_k \models \phi.$

\subsection{Fragments and compilation between logics}

We define a \emph{primitive positive formula} to be a formula built from
atoms, conjunction, and existential quantification.
We define an \emph{existential-positive formula} to be a formula built from
atoms, conjunction, disjunction, and existential quantification.
We define a \emph{positive first-order formula} to be a formula that does not
use negation.  
We refer to an existential-positive formula that is a sentence as an
\emph{existential-positive sentence}, and extend this convention to define
primitive positive sentences, positive first-order sentences, etc. correspondingly.

We introduce the following notation, for each $k \geq 1$.
\begin{itemize}

\item Let $\PP{k}$ denote the class containing each primitive positive sentence
that uses at most $k$ variables.

\item Let $\exPP{k}$ denote the class containing each primitive positive sentence
that is logically equivalent to a sentence in $\PP{k}$.

\item Let $\EP{k}$ denote the class containing each existential-positive sentence
that uses at most $k$ variables.

\item Let $\exEP{k}$ denote the class containing each existential-positive sentence
that is logically equivalent to a sentence in $\EP{k}$.

\item Let $\PFO{k}$ denote the class containing each positive first-order sentence
that uses at most $k$ variables.

\end{itemize}
By a sentence that \emph{uses at most $k$ variables}, we mean that when all variables
occurring in the sentence are placed into a set, that set has size $k$ or less.
For each of these fragments, when $\sigma$ is a signature, we add $\sigma$ as a subscript
to restrict to sentences over $\sigma$, so for example, $\EP{k}_\sigma$
denotes the class of all sentences in $\EP{k}$ that are over signature $\sigma$.

\begin{prop} 
\label{prop:unarycompilestoEPone}
When $\sigma$ is a signature that consists
only of propositional symbols and unary symbols,
it holds that every existential-positive sentence
over $\sigma$ is in $\exEP{1}_\sigma$.
\end{prop}

\begin{proof}
This
follows directly from the argument in the first paragraph
of~\cite[Proof of Theorem 4.3]{Chen-tocl}.
\end{proof}

The \emph{size} of a formula $\psi$, denoted by $\| \psi \|$, is defined inductively:
$\| \alpha \| = 1$ when $\alpha$ is an atom;
$\| \neg \phi \| = \| \phi \| + 1$; 
$\| \phi \vee \psi \|$ and $\| \phi \wedge \psi \|$ are defined as
$\| \phi \| + \| \psi \| + 1$;
and $\| \exists v \phi \|$ and $\| \forall v \phi \|$
are defined as $\| \phi \| + 1$.
So, the size of a formula is essentially the number of atom occurrences plus the number of
occurrences of connectives and quantifiers.

When $\fancyl$ and $\fancyl'$ are logics,
a \emph{compilation} from
$\fancyl$ to $\fancyl'$ is a mapping
$c: \fancyl \to \fancyl'$
such that for each $\phi \in \fancyl$,
it holds that $\phi$ and $c(\phi)$ are logically equivalent.
The \emph{size} of a compilation $c$ is 
the function $f: \N \to \N$ defined by
$$f(n) = \max \{ \| c(\phi) \| ~|~ \phi \in \fancyl, \| \phi \| = n \}.$$

\subsection{Parameterized complexity and compilation}
\label{subsect:parameter-compilation}

We present the framework that we will use for ruling out compilations to 
arbitrary polynomial-size formats.  
The framework given here comes from work~\cite{pc-ipec,pc-arxiv} of the second author.

Here, we use $\ptime$ to denote the class of polynomial-time computable languages,
and $\fptime$ to denote the class of polynomial-time computable functions.

By a \emph{parameterization}, we refer
to a polynomial-time computable map from $\Sigma^*$ to $\Sigma^*$.
Relative to a parameterization $\kappa: \Sigma^* \to \Sigma^*$,
we say that
$\kappa(x)$
is
the \emph{parameter}  
of the string $x$.
It is typical in the literature to define
a parameterization to be a map from $\Sigma^*$ to $\N$;
however,
in the present article we want to apply 
compilation functions to parameters represented as strings and
discuss the \emph{length} of the results.
In particular, we will study query evaluation, and the parameter
of each instance will be the query to be evaluated.

A \emph{parameterized problem} is a pair $(Q, \kappa)$
consisting of a language $Q$ and a parameterization $\kappa$.
The parameterized problems that will be of focal study in this article are defined as follows.
\begin{definition}
When $\fancyl$ is a logic, we define $\qe(\fancyl)$ to be the parameterized problem
$(Q,\kappa)$ whose language $Q$ is 
$$\{ (\phi, \relb) ~|~ \phi \in \fancyl, \relb \models \phi \}$$
and where $\kappa(\phi,\relb) = \phi$.
\end{definition}

A function $c: \Sigma^* \to \Sigma^*$ is said to be
\emph{polynomial length} if there exists a polynomial $\ell: \N \to \N$
such that
for each $x \in \Sigma^*$,
it holds that $|c(x)| \leq \ell(|x|)$.
A function $g: \Sigma^* \to \Sigma^*$ is
\emph{poly-compilable} 
with respect to a parameterization $\kappa$
if there exist $f \in \fptime$ and
a computable, polynomial length function $c: \Sigma^* \to \Sigma^*$
such that (for each $x \in \Sigma^*$) it holds that
$g(x) = f(c(\kappa(x)),x)$.

We now present the parameterized complexity classes that will be used.
Let us say that 
a parameterized problem $(Q, \kappa)$ 
is \emph{poly-compilable to $\fancyc$}
if
there exists a  function $g: \Sigma^* \to \Sigma^*$
that is poly-compilable (with respect to $\kappa$)
and a language $Q' \in \fancyc$
such that (for each $x \in \Sigma^*$)
$x \in Q \Leftrightarrow g(x) \in Q'$.
We define $\polycompp$ 
to be the 
class of parameterized problems that contains each parameterized problem that
is poly-compilable to $\ptime$;
and, we define $\polycomp{\np}$
to be the 
class of parameterized problems that contains each parameterized problem that
is poly-compilable to $\np$.
We define $\chopped{\np}$ to be the class of parameterized problems
that contains each parameterized problem $(Q, \kappa)$
that is in $\polycomp{\np}$ via a poly-compilable function $g$
for which there exists a polynomial $p: \N \to \N$
such that (for each $x \in \Sigma^*$) $|g(x)| \leq p(|\kappa(x)|)$.

We will use the following notion of reduction.
We say that a parameterized problem $(Q, \kappa)$
\emph{poly-comp reduces}
to another parameterized problem $(Q', \kappa')$
if 
there exists a function $g: \Sigma^* \to \Sigma^*$
that is poly-compilable with respect to $\kappa$
and a 
poly-length, computable function $s: \Sigma^* \to \wpfin(\Sigma^*)$
such that (for each $x \in \Sigma^*$)
it holds that $x \in Q \Leftrightarrow g(x) \in Q'$
and that $\kappa'(g(x)) \in s(\kappa(x))$.
Here, when $T$ is a set, we use $\wpfin(T)$ to denote the power set of $T$
restricted to finite sets, that is, the
set $\{ S \subseteq T ~|~ S \textup{ is finite} \}$.
We consider a parameterized problem to be 
$\chopped{\np}$-hard if every problem in $\chopped{\np}$ 
poly-comp reduces to it.

\begin{theorem} \cite[Theorem 13]{pc-ipec,pc-arxiv}
\label{thm:poly-comp-red-transitive}
Poly-comp reducibility is transitive.
\end{theorem}

We will establish $\chopped{\np}$-hardness results which provide evidence
for non-containment in the parameterized class $\polycompp$, as the following theorem shows.

\begin{theorem} (follows directly from~\cite[Proposition 9 and Theorem 19]{pc-ipec,pc-arxiv})
\label{thm:chopped-np-not-in-polycompp}
A problem that is $\chopped{\np}$-hard is not contained in the 
class $\polycompp$,
unless the polynomial hierarchy collapses.
\end{theorem}

In order to establish $\chopped{\np}$-hardness results, we will give reductions
from the following problem.
Let $\clique$ denote the clique problem, that is, the set of pairs $(G,k)$ 
such that $G$ is an undirected graph, $k \geq 1$, and $G$ contains a clique of size $k$.
Define $\nuclique$ to be the parameterized problem $(\clique, \nu)$
where $\nu$ is the parameterization such that $\nu(G,k)$ is equal to the number
of vertices in $G$, expressed in unary; that is, $\nu(G,k) = 1^n$, where $n$ is the
number of vertices in $G$.

\newcommand{\len}{\mathrm{len}}

\begin{theorem} 
\label{thm:nuclique-chopped-np-hard}
The parameterized problem $\nuclique$ is $\chopped{\np}$-hard.
\end{theorem}

\begin{proof}
By the argument in \cite[Proof of Theorem 30]{pc-arxiv}, 
the parameterized problems $\nuclique$ and $(\clique, \len)$ are poly-comp interreducible;
here, $\len$ denotes the parameterization that, given a string, returns the string's length
in unary (that is, given $x$, returns $1^{|x|}$).
The parameterized problem $(\clique, \len)$ is $\chopped{\np}$-hard 
by \cite[Proposition 16]{pc-ipec,pc-arxiv}.
\end{proof}

\section{Upper Bounds}

\subsection{General upper bound}

We establish the following general single-exponential upper bound on the length of compiling
an existential-positive formula to the bounded-variable fragment.

\begin{theorem} (General upper bound)
\label{thm:upper-bound-general}
There exists a polynomial $q$ such that
for each $k \geq 1$,
there is a compilation $c$ from $\exEP{k}$ to $\EP{k}$
behaving as follows:
when $\phi \in \exEP{k}$ is a sentence
with $\| \phi \| = n$,
it holds that $\| c(\phi) \| \leq 2^n q(n)$.
\end{theorem}

To establish Theorem~\ref{thm:upper-bound-general},
we in fact prove the following theorem, which provides a more refined upper bound
in the case of prenex sentences.

\begin{theorem} (General upper bound on prenex sentences)
\label{thm:upper-bound-in-terms-of-atoms}
There exists a polynomial $q$ such that
for each $k \geq 1$,
there is a compilation $c$ from $\exEP{k}$ to $\EP{k}$
behaving as follows:
when $\phi \in \exEP{k}$ is a sentence in prenex normal form
with $m$ occurrences of non-propositional atoms
and $\| \phi \| = n$,
it holds that $\| c(\phi) \| \leq 2^m q(n)$.
\end{theorem}

Theorem~\ref{thm:upper-bound-in-terms-of-atoms} implies Theorem~\ref{thm:upper-bound-general}:
any first-order formula can converted to prenex normal form
via syntactic transformations that do not change its size;
and, for any first-order formula,
the number of occurrences of non-propositional atoms is bounded above by the formula's size.

Before proving Theorem~\ref{thm:upper-bound-in-terms-of-atoms}, we establish a lemma.
For any set of first-order formulas $\Phi$, we let
$\real(\Phi)$  be the set of all $\Phi' \subseteq \Phi$ such that there is an
  interpretation $I$ with  $I\models \psi \Longleftrightarrow
  \psi\in\Phi'$ for all $\psi \in \Phi$.

\begin{lemma}
\label{lemma:disjunctive-form}
There exists a polynomial $p$ such
  that,
when $\varphi$ is a 
formula that is a positive Boolean combination 
of propositional atoms and
of positive formulas drawn from a finite set $\Phi$,
   the formula $\varphi$ is logically equivalent to a
   formula $\varphi'$ such that
   $$\|\varphi'\| \leq |\real(\Phi)|\cdot p(\|\varphi\|),$$
  and where $\varphi'$ has the form
  $$\bigvee_{\Phi'\in\real(\Phi)} \bigl(\varphi_{\Phi'} \wedge
  \bigwedge_{\psi\in\Phi'}\psi\bigr),$$ 
  where 
  each $\varphi_{\Phi'}$ is a
  positive Boolean combination of propositional atoms of size $O(\|\varphi\|)$.
\end{lemma}

\begin{proof}
  For any $\Phi' \in  \real(\Phi)$ let $\varphi_{\Phi'}$ be the formula
  obtained from $\varphi$ by replacing every subformula in $\Phi'$
  with $\top$ and every subformula in $\Phi\setminus \Phi'$ with
  $\bot$.
  We need to show that $\varphi$ and $\varphi'$ are equivalent. Let
  $I$ be an interpretation that satisfies $\varphi$ and let
  $\Phi'\subseteq \Phi$ be the set of formulas satisfied by $I$. It
  follows from the definition that $I$ satisfies $\varphi_{\Phi'} \wedge
  \bigwedge_{\psi\in\Phi'}\psi$ and hence $\varphi'$. For the other
  direction assume that $I$ satisfies $\varphi'$ and let
  $\Phi'\subseteq \Phi$ be such that $I$ satisfies the disjunct $\varphi_{\Phi'} \wedge
  \bigwedge_{\psi\in\Phi'}\psi$. Since $\varphi$ is a positive in the stated sense, it
  follows that $I\models \varphi_{\Phi''}$ where
  $\Phi''\supseteq\Phi'$ contains exactly those formulas from $\Phi$
  that are satisfied by $I$. By the definition of $\varphi_{\Phi''}$ it therefore holds that $I\models\varphi$.  
\end{proof}

Let us remark that the just-given lemma is effective when the set $\Phi$
consists of primitive positive formulas, in the sense that $\varphi'$
can be computed from $\varphi$.
This is because entailment on primitive positive formulas
is decidable, so the set $\real(\Phi)$ can be computed
from $\Phi$.

\begin{proof}[Proof of Theorem~\ref{thm:upper-bound-in-terms-of-atoms}]
  The compilation $c$ is described as follows.  
Let
  $\phi = \exists x_1 \ldots \exists x_t \phi^-$
  be an
  $\exEP{k}$-sentence 
  in prenex normal form and let $\Phi$ be the set
  whose elements are all
  non-propositional atoms that occur in $\phi^-$.  
We have $|\Phi| \leq m$, so 
$|\real(\Phi)| \leq 2^m$.
Apply Lemma~\ref{lemma:disjunctive-form}
to the quantifier-free part $\phi^-$ of $\phi$; 
one obtains a sentence of the form
$$\exists x_1 \ldots \exists x_t (\bigvee_{i=1}^r \phi^i)$$
and having size at most $2^m \cdot p(||\phi||)$,
where $p$ is the polynomial from
Lemma~\ref{lemma:disjunctive-form}; 
this sentence is logically equivalent to
$\bigvee_{i=1}^r \exists x_1 \ldots \exists x_t \phi^i$.
Here, due to our choice of $\Phi$
and
the statement of Lemma~\ref{lemma:disjunctive-form},
each $\phi^i$ is a conjunction that contains (as a conjunct) up to one 
positive Boolean combination of propositional atoms, and otherwise
contains non-propositional atoms.
Each such $\phi^i$ can be rewritten, in a way that preserves logical equivalence, 
so that it contains no equalities;
this is done by taking an equality $u = v$, 
replacing each instance of $v$ with $u$,
and then deleting the equality.

Now rewrite this last sentence so that it is
\emph{irredundant} in the sense that
for no distinct $i,j$ does it hold that
$\exists x_1 \ldots \exists x_t \phi^i$ 
entails $\exists x_1 \ldots \exists x_t \phi^j$; 
this can be done simply
by removing any disjunct 
$\exists x_1 \ldots \exists x_t \phi^i$ such that 
the entailment holds.
Now, it is known from~\cite[Lemma 4]{BovaChen14}
and from the assumption that $\phi \in \exEP{k}$
that
each such disjunct of the irredundant sentence
(neglecting the possible conjunct that is a positive Boolean of propositional atoms)
is in $\exPP{k}$;  each sentence
in $\exPP{k}$ can be converted to 
a sentence in $\PP{k}$ without increasing its size~\cite{DKV}.
This yields the theorem, for
ever since the bound 
of $2^m \cdot p(||\phi||)$ was computed,
the transformations that have been applied
were to duplicate the quantifier prefix,
remove disjuncts, and perform the conversions
from $\exPP{k}$ to $\PP{k}$.
\end{proof}

\subsection{The case of $1$ variable}

Theorem~\ref{thm:upper-bound-in-terms-of-atoms}
gave a general exponential upper bound on the size of
a compilation from $\exEP{k}$ to $\EP{k}$.
Now we turn to the 1-variable case. We provide two upper bounds: the
first one in the case where the signature $\sigma$ is finite and the
second when $\sigma$ consists of at
most one non-propositional symbol, and any number of propositional symbols
(even infinitely many).

\begin{lemma}
\label{lem:upper-bound-for-one-variable-fin-symbols}
Let $\sigma$ be a finite signature.
There exists
a compilation $c$ from $\exEP{1}$ to $\EP{1}$
with $\| c(\phi) \| = O(2^{2^{\|\sigma\|}})$.
\end{lemma}

\begin{proof}
We first make an observation.  For any structure
$\relb$, define $\relb^-$ to be the structure
with the same universe as $\relb$
and where 
$$R^{\relb^-} = 
R^{\relb} \cap \{ (b, \ldots, b) ~|~ b \in B \}$$
for each relation symbol $R$.
Then, when $\psi \in \exEP{1}$, it holds that
$\relb \models \psi$ if and only if
$\relb^- \models \psi$.
It is sufficient to argue this 
for all $\psi \in \EP{1}$; this holds
because each atom $R(v_1,\ldots,v_k)$ in such a $\psi$
has $v_1 = \cdots = v_k$.

We next turn to prove the theorem.
Let $\phi \in \exEP{1}$ be as described,
and let us assume that $\phi$ is in prenex form.
Define $\phi^=$ to be the rewriting of $\phi$
where each atom $R(v_1,\ldots,v_k)$ is replaced with
$$R(v_1, \ldots, v_1) \wedge (v_1 = v_2) \wedge
(v_2 = v_3) \wedge \cdots \wedge (v_{k-1} = v_k).$$
We have, for any structure $\relb$:
$$\relb \models \phi 
\Leftrightarrow \relb^- \models \phi
\Leftrightarrow \relb^- \models \phi^=
\Leftrightarrow \relb \models \phi^=
.$$
Here, the first and last equivalences follow from
the observation.
We conclude that $\phi$ and $\phi^=$
are logically equivalent.

Let $\exists x_1 \ldots \exists x_t$ 
be the quantifier prefix
of $\phi$ and $\phi^=$,
and let $R_1, \ldots, R_s$ be the non-propositional
symbols that appear in $\phi$ and $\phi^=$.
Observe that the quantifier-free part of $\phi^=$
is a positive Boolean combination of propositional atoms, atomic formulas $R(x_i,\ldots,x_i)$ and
equalities $x_i = x_j$. 

First, we transform this quantifier free part to DNF form.
After doing so, the quantification
$\exists x_1 \ldots \exists x_t$ may be pushed in
so that the result is a disjunction $\phi^\vee$
of formulas each having the form
$\exists x_1 \ldots \exists x_t \bigwedge_{\psi \in \Phi'} \psi$;
let us now refer here to such a formula as a disjunct.

Consider each disjunct
$\exists x_1 \ldots \exists x_t \bigwedge_{\psi \in \Phi} \psi$.
One may eliminate each variable equality 
$x_i = x_j$ by replacing every instance of $x_j$
with $x_i$.  Then, by grouping together conjuncts
that refer to the same variable, 
it can be seen that
such a formula is logically equivalent to
a conjunction of formulas each having the form
$\theta_\tau = 
\exists x \bigwedge_{R \in \tau} R(x,\ldots,x)$,
where $\tau \subseteq \sigma$.
Overall, we get a formula that is a disjunction of conjunctions of
formulas $\theta_\tau$. It follows that the size of the formula is bounded by $2^{2^{O(\|\sigma\|)}}$.
\end{proof}

\begin{lemma}
\label{lem:upper-bound-for-one-variable-one-nonprop-symbol}
Let $\sigma$ be a signature containing at most one non-propositional symbol.
There exists
a compilation $c$ from $\exEP{1}$ to $\EP{1}$
with $\| c(\phi) \| = \|\phi\|$.
\end{lemma}

\begin{proof}
  The compilation is fairly simple: we just replace every variable in
  $\phi$ by $x$.
  It remains to prove that this is correct. If $\sigma$ contains only
  propositional symbols, then the lemma is trivial. Otherwise, let $ R$ be the
  non-propositional symbol in $\sigma$ and $\relb$ an
  arbitrary $\sigma$-structure. We define the $\sigma$-structure
  $\relb^-$ over the singleton universe $B^-=\{1\}$ by interpreting the propositional symbols as in
  $\relb$ and defining $ R^{\relb^-}=\{(1,\ldots,1)\}$ if there
  is an $b\in B$ with $(b,\ldots,b)\in R^{\relb}$ and
  $ R^{\relb^-}=\emptyset$ otherwise.
  It is not hard to verify that for every $\psi\in\EP{1}$ it holds
  that $\relb\models \psi \,\Leftrightarrow\,\relb^-\models \psi$ and
  hence the same holds true for every $\psi\in\exEP{1}$. Moreover,
  since $\relb^-$ is a structure with singleton universe, it holds
  that $\relb^-\models \phi\,\Leftrightarrow\,\relb^-\models
  c(\phi)$. Thus, for any structure $\relb$ we have
  $$
   \relb\models \phi
   \,\Leftrightarrow\,
   \relb^-\models \phi
   \,\Leftrightarrow\,
   \relb^-\models c(\phi)
   \,\Leftrightarrow\,
   \relb\models c(\phi),
  $$
  showing that $\phi$ and $c(\phi)$ are logically equivalent.
\end{proof}

\section{Classification theorems}
\label{sec:classificationtheorems}

We state our classification theorems, and give proofs that make forward references
to the next section, which contain the lower bounds; in particular,
these proofs contain forward references to Theorems~\ref{thm:lower-bound-inf-prop-two-unary}
and
\ref{thm:lower-bound-at-least-binary}.

\begin{theorem}\label{thm:classification-one-variable}
Let $\sigma$ be a signature.
\begin{itemize}
\item 
If $\sigma$ is finite or contains at most one non-propositional symbol, then there exists 
a polynomial-size compilation from 
$\exEP{1}_\sigma$
to $\EP{1}_\sigma$.
\item
Otherwise, 
for all $\ell \geq 1$, 
there exists $\epsilon > 0$ such that
any compilation from
$\exEP{1}_\sigma$
to $\PFO{\ell}_\sigma$ 
has size $\Omega(2^{n^{\epsilon}})$.
\end{itemize}
\end{theorem}

\begin{proof}
The upper bound follows from Lemma~\ref{lem:upper-bound-for-one-variable-fin-symbols} 
and Lemma~\ref{lem:upper-bound-for-one-variable-one-nonprop-symbol}.
The lower bound follows from 
Theorem~\ref{thm:lower-bound-inf-prop-two-unary}.
\end{proof}

\begin{theorem}
Let $k \geq 2$, and let $\sigma$ be a signature.
\begin{itemize}
\item 
If $\sigma$ contains only propositional and unary symbols and,
in addition, is either finite or contains at most one unary symbol, 
then there exists a polynomial-size compilation
from $\exEP{k}_\sigma$ to $\EP{k}_\sigma$.
\item
Otherwise,
for all $\ell \geq k$, 
there exists $\epsilon > 0$ such that
any compilation from
$\exEP{k}_\sigma$
to $\PFO{\ell}_\sigma$ 
has size $\Omega(2^{n^{\epsilon}})$.
\end{itemize}
\end{theorem}

\begin{proof}
If the signature contains no symbols of arity strictly larger than one, then by
Proposition~\ref{prop:unarycompilestoEPone} we have 
that $\exEP{k}_\sigma$
and $\exEP{1}_\sigma$ are equal. 
Hence, in this case the upper as well as the lower bounds immediately follow from Theorem~\ref{thm:classification-one-variable}.
If $\sigma$ contains one at
least binary symbol, then the lower bound follows from
Theorem~\ref{thm:lower-bound-at-least-binary}.
\end{proof}

\section{Lower bounds}

On a high level,
our unconditional lower bounds on compilations to bounded-variable positive logics
will be based on previously established circuit lower bounds in conjunction with
the fact (given soon as Lemma~\ref{lem:formulas-to-circuits})
that formulas can be translated to circuits that, in a way to be made precise,
preserves formula semantics.
As we establish these unconditional lower bounds, we also establish hardness results
in the parameter compilation framework presented in 
Section~\ref{subsect:parameter-compilation};
in particular, we establish
$\chopped{\np}$-hardness results.

A monotone Boolean circuit $C_N$ of size $m$ is defined to be
a directed acyclic graph
over $m$ vertices (the gates) with a designated
output gate, $N$ sources (the inputs), and where every
non-source vertex is labelled with either $\land$ or $\lor$.
The circuit naturally computes a monotone function $C_N\colon \{0,1\}^N\to\{0,1\}$.
Assume $\sigma$ to be a finite signature and $n \geq 1$;
a monotone Boolean circuit $C^\sigma_n$ that computes a function over
all $\sigma$-structures $\rela$ with universe $A=[n]$ 
is a monotone Boolean circuit $C_N$ that contains, 
for every pair consisting of an $r$-ary relation symbol $R$ and an
$r$-tuple $(v_1,\ldots,v_r) \in [n]^r$,
one input gate
that indicates whether
$(v_1,\ldots,v_r)\in R^\rela$. 
We consider the \emph{size} of a monotone Boolean circuit to be the number
of gates that it contains.

In what follows, we use the term \emph{$k$-ary signature} to refer to a signature wherein all
relation symbols have arity $k$ or less.

\begin{lemma} (from formulas to circuits)
\label{lem:formulas-to-circuits}
For all $n,k\geq 1$, finite $k$-ary
signatures $\sigma$ and
$\varphi\in\PFO{k}_\sigma$ there exists
a monotone circuit $C^\sigma_n$ of size at most $(\|\varphi\|+\|\sigma\|)n^k$ that receives a
$\sigma$-structure $\rela$ with universe $A=[n]$ as input and decides whether $\rela\models \varphi$. 
\end{lemma}

\begin{proof}
  The monotone circuit contains for every subformula
  $\psi(x_1,\ldots,x_k)$ of $\varphi$ and every tuple $\tupv=(v_1,\ldots,v_k)\in
  [n]^k$ a gate $g^\psi_{\tupv}$ that computes whether
  $\rela,\tupv\models \psi$. The construction of the circuit follows
  the inductive definition of $\varphi$: 
  \begin{itemize}

  \item 
  If $\psi=R(x'_1,\ldots,x'_r)$ is an atom, then $g_{\psi,\tupv}$ is 
  defined to be the input gate for the pair
  consisting of $R$ and
  the arity $r$ subtuple $\tupv'\subseteq \tupv$ corresponding to
  $x'_1, \ldots, x'_r$.

  \item
  If $\psi=\psi_1\ast\psi_2$ for
  $\ast\in\{\land,\lor\}$, then $g_{\psi,\tupv}$ is an $\ast$-gate
  with inputs $g_{\psi_1,\tupv}$ and $g_{\psi_2,\tupv}$.

  \item Finally, we let $\psi=\exists x_i \psi'$ ($\psi=\forall x_i
    \psi'$) and denote by $\tupv_w$ the $k$-tuple $(v_1,\ldots,v_{i-1},w,v_{i+1},\ldots,v_k)$.
    The gate
    $g_{\psi,\tupv}$ is an $\lor$-gate ($\land$-gate) with $n$ inputs
    $g_{\psi',\tupv_w}$ for $w\in[n]$. 
  \end{itemize}
  It is straightforward to verify that (for any $\tupv$) the gate $g_{\varphi,\tupv}$
  decides whether $\rela\models\varphi$.
  By construction, the circuit has
  $\sum_{R\in\sigma}n^{\ar(R)}$ input gates and at most $\|\varphi\|n^k$ non-input  gates.
\end{proof}

We will make use of the following size lower bound for monotone circuits computing the
clique function.

\begin{theorem}[\cite{AlonBoppana87}]
  \label{thm:monotone-clique}
  Let $\sigma=\{E\}$ be the signature of graphs and $k(n) = \bigl\lfloor
  \tfrac14(n/\log n)^{2/3}\bigr\rfloor$.
  The size of monotone circuits $C^\sigma_n$ that decide $k(n)$-clique on $n$-vertex graphs is lower bounded by $2^{\Omega((n/\log n)^{1/3})}$.
\end{theorem}

\subsection{Infinite signatures with two distinct non-propositional symbols}

\begin{theorem}
\label{thm:lower-bound-inf-prop-two-unary}
If $\sigma$ is an infinite signature that contains two distinct non-propositional symbols, then for each $\ell \geq 1$,
any compilation from
$\exEP{1}_\sigma$ to $\PFO{\ell}_\sigma$
has size
$\Omega(2^{\formsizeN^{1/14}})$.
\end{theorem}

\begin{proof}
We prove this theorem under the assumption that $\sigma$ contains infinitely many propositional symbols.  This suffices, since when $R$ is a non-propositional symbol of arity $r$,
we may use the formula $\exists x R(x, \ldots, x)$ to simulate a propositional atom.
We also make the assumption that $\sigma$ contains two distinct unary symbols $S_0$ and $S_1$;
this is justified, for if $T_0$ and $T_1$ are distinct non-propositional symbols,
one may use $T_0(v, \ldots, v)$ and $T_1(v, \ldots ,v)$ 
to simulate the atoms $S_0(v)$ and $S_1(v)$, respectively.

  For $i,j\in\N$ we let $P_{ij}$ be a propositional symbol in $\sigma$
  and use $S_0$ and $S_1$ to denote two distinct unary symbols in $\sigma$. 
  Let
  $$\sigma_n=\setc{P_{ij}}{i,j\in[n]}\cup\{S_0,S_1\}.$$
  We have that $\sigma_n \subseteq \sigma$.

  Suppose that $\graphG$ is a graph on $n=2^d$ vertices $v_1,\ldots,v_{n}$ for
  some $d\in\N$.
  We let $\rela_\graphG$ be the
  $\sigma$-structure with universe $A_\graphG=\{0,1\}$,
  $S_0^\rela=\{0\}$, $S_1^\rela=\{1\}$,   
  \begin{equation}
    \label{eq:1} 
  P^\rela_{ij} =
  \begin{cases}
    \{\emptyset\}\text{, if }(v_i,v_j)\in E^\graphG \\
    \emptyset\text{, else,}
  \end{cases}
  \end{equation}
  for all $i,j\in[n]$, and $R^\rela=\emptyset$ for all $R\in
  \sigma\setminus \sigma_n$.
  \newcommand{\binenc}[1]{\ensuremath{\langle #1\rangle}}

  Now we define an $\EP{k}_\sigma$ sentence $\varphi_{n,k}$,
  which is satisfied by $\rela_\graphG$ if and only if $\graphG$ contains a
  $k$-clique. For every $i\in[n]$ we write
  $\binenc{i}=(b_1,\ldots,b_d)\in\{0,1\}^d$ to denote the $d$-bit binary
  encoding of the number $i-1$.
  We define
  \begin{align*}
    \label{eq:2}
    \varphi_{n,k} &= \exists x^1_1 \,\cdots\, \exists x^d_1\quad \exists x^1_2
    \,\cdots\, \exists x^d_2 \quad \cdots\quad \exists x^d_k \\
    &\!\!\!\!\!\!\!\!
    \bigwedge_{\substack{\\i,j\in [n]\\s,t \in [k]\\s \neq t}}
    \!
    \bigl(P_{ij}
    \vee
    \!\!\!\!\!\!
    \bigvee_{\substack{(b_1,\ldots,b_d)\\\in \{0,1\}^d \setminus \{\binenc{i}\}}}
    \bigwedge_{r\in[d]}
    S_{b_r}(x^r_s)
    \vee
    \!\!\!\!\!\!
    \bigvee_{\substack{(b_1,\ldots,b_d)\\\in \{0,1\}^d \setminus \{\binenc{j}\}}}
    \bigwedge_{r\in[d]}
    S_{b_r}(x^r_t)
    \bigr).
  \end{align*}
The existential quantifiers guess the bit encoding of $k$ vertices
and the quantifier-free part ensures that they form a clique. 
In particular, the subformula
$$    \bigvee_{\substack{(b_1,\ldots,b_d)\\\in \{0,1\}^d \setminus \{\binenc{i}\}}}
    \bigwedge_{r\in[d]}
    S_{b_r}(x^r_s)$$
is true if and only if the $s$th guessed vertex is not $i$;
precisely, it is true if and only if the bits
$x_s^1, \ldots, x_s^d$ are not equal to $\binenc{i}$.
Analogously, the subformula 
$$    \bigvee_{\substack{(b_1,\ldots,b_d)\\\in \{0,1\}^d \setminus \{\binenc{j}\}}}
    \bigwedge_{r\in[d]}
    S_{b_r}(x^r_t)$$
is true if and only if the $t$th guessed vertex is not $j$;
precisely, it is true if and only if the bits
$x_t^1, \ldots, x_t^d$ are not equal to $\binenc{j}$.
Hence, the quantifier-free part of $\varphi_{n,k}$ is true if and only if
for all distinct $s, t \in [k]$, 
letting $i'$ and $j'$ denote the $s$th and $t$th guessed vertices,
$P_{i'j'}$ holds; that is, if and only if
the guessed vertices form a clique in $\graphG$.

We have that $\varphi_{n,k}\in \exEP{1}_\sigma$ by
Proposition~\ref{prop:unarycompilestoEPone}. Let $\psi_{n,k}$ be an
$\PFO{\ell}_\sigma$ sentence that is equivalent to $\varphi_{n,k}$.
By Lemma~\ref{lem:formulas-to-circuits} there exists a monotone Boolean circuit
$C^\sigma_{n}$ having size $O(\|\psi_{n,k}\|n^\ell)$ that evaluates $\psi_{n,k}$ on $\sigma_n$-structures over an 
$n$-element universe. By hard-coding all inputs that are different
from $F_{ij}$, we obtain a monotone circuit $C^{\{E\}}_{n}$ that receives the
adjacency matrix of an $n$-vertex graph $\graphG$ and decides whether
$\rela_\graphG$ satisfies $\psi_{n,k}$. Since this is the case if and
only if $\graphG$ contains a $k$-clique, we can choose $k=\bigl\lfloor
  \tfrac14(n/\log n)^{2/3}\bigr\rfloor$ and obtain from
  Theorem~\ref{thm:monotone-clique} that $C^\sigma_{n}$ has size $2^{\Omega((n/\log n)^{1/3})}$.
It follows that $\|\psi_{n,k}\|$ is lower bounded by
$2^{\Omega((n/\log n)^{1/3}-\ell\log n)}$. The Theorem follows from the observation that 
$\formsizeN=\|\varphi\|=O(k^2n^3\log n)=O(n^{13/3})$. 
\end{proof}

\begin{theorem}
\label{thm:chopped-np-two-distinct-non-prop}
When $\sigma$ is an infinite signature that contains two distinct non-propositional symbols,
the parameterized problem 
$$\qe(\exEP{1}_\sigma)$$
is $\chopped{\np}$-hard.
\end{theorem}

\begin{proof}
We present a poly-comp reduction $(g,s)$ from the problem $\nuclique$;
this suffices by Theorem~\ref{thm:nuclique-chopped-np-hard}
and Theorem~\ref{thm:poly-comp-red-transitive}.
The mapping $g$, given an instance $(\graphG,k)$ of $\clique$,
outputs the pair $(\phi_{n,k},\rela_\graphG)$ as defined 
in the proof of 
Theorem~\ref{thm:lower-bound-inf-prop-two-unary}; 
here, $n$ is the number of vertices in $G$.
(We define $g$ only on well-formed instances, in particular,
instances with $k \leq n$; it is straightforwardly verified that this suffices.)
This mapping is clearly polynomial-time computable;
also, by the argumentation 
in the proof of 
Theorem~\ref{thm:lower-bound-inf-prop-two-unary},
it holds that $(\graphG,k)$ is a yes instance of $\clique$
if and only if 
$\rela_\graphG \models \phi_{n,k}$
if and only if 
$(\phi_{n,k},\rela_\graphG)$ is a yes instance of
$\qe(\exEP{1}_\sigma)$.
The mapping $s$ is defined by 
$s(n) = \{ \phi_{n,i} ~|~ i \leq n \}$.
For all instances $(G,k)$,
we then have that $\phi_{n,k} \in s(n)$.
\end{proof}

\subsection{Signatures having an at least binary symbol}

\begin{theorem}
\label{thm:lower-bound-at-least-binary}
If a signature $\sigma$ contains 
an at least binary symbol, 
then for each $\ell \geq 2$,
there exists $\epsilon > 0$ such that
any compilation from
$\exEP{2}_\sigma$ to $\PFO{\ell}_\sigma$
has size 
$\Omega(2^{n^{\epsilon}})$.
\end{theorem}

We let $E$ be a binary symbol, and prove the theorem
in the case that $\sigma = \{ E \}$; this suffices, since a higher arity symbol
can simulate a binary symbol.

For each $n \geq 1$ and each $i \in [n]$,
we define a structure $\reld_{i,n}$ as follows.
The universe $D_{i,n}$ is equal to 
$$\{ d \} \cup \{ p_i^1, \ldots, p_i^i \} \cup 
            \{ q_i^{i-1}, q_i^i, \ldots, q_i^{n+1} \},$$
and $E^{\reld_{i,n}}$ is defined as
$$\{ (d, p_i^1) \} \cup 
 \{ (p_i^k, p_i^{k+1}) ~|~ k = 1, \ldots, i-1 \} \hspace{2pt}
\cup $$
$$ \{ (q_i^{i-1}, p_i^i) \}
 \cup
 \{ (q_i^k, q_i^{k+1}) ~|~ k = i-1, \ldots, n \}.
$$
We refer the reader to the figure for a visualization of this structure.

Observe that for distinct values $i, i' \in [n]$,
the only element in both of the universes 
$D_{i,n}$, $D_{i',n}$ is $d$, that is, we have
$D_{i,n} \cap D_{i',n} = \{ d \}$.
For a non-empty subset $I \subseteq [n]$,
let us use $\reld_{I,n}$
to denote $\cup_{i \in I} \reld_{i,n}$.
In what follows, we will overload $\N$,
and use it to denote the structure whose
universe is the set $\N$ of natural numbers
and where $E^{\N} = \{ (k, k+1) ~|~ k \in \N \}$.

\begin{figure}
      \centering
\begin{tikzpicture}
  [vertex/.style={circle,draw=black,fill=black,inner sep=0pt,minimum  size=3 pt}]
  \node[vertex,label={$d$}] (1) at (0,0) {};
  \node[vertex,label={$p_i^1$}] (2) at (.75,0) {};
  \node[vertex,label={$p_i^2$}] (3) at (1.5,0) {};
  \node[vertex,label={$p_i^{i-1}$}] (4) at (2.5,0) {};
  \node[vertex,label={$p_i^i$}] (5) at (3.25,0) {};
  \node[vertex,label=below:{$q_i^{i-1}$}] (6) at (2.5,-.5) {};
  \node[vertex,label=below:{$q_i^i$}] (7) at (3.25,-.5) {};
  \node[vertex,label=below:{$q_i^{i+1}$}] (8) at (4,-.5) {};
  \node[vertex,label=below:{$q_i^{n-1}$}] (9) at (5,-.5) {};
  \node[vertex,label=below:{$q_i^{n}$}] (10) at (5.75,-.5) {};
  \draw[-latex,shorten >= 0.5mm,shorten <= 0.5mm] (1) -- (2);
  \draw[-latex,shorten >= 0.5mm,shorten <= 0.5mm] (2) -- (3);
  \draw[-latex,shorten >= 0.5mm,shorten <= 0.5mm,dotted] (3) -- (4);
  \draw[-latex,shorten >= 0.5mm,shorten <= 0.5mm] (4) -- (5);
  \draw[-latex,shorten >= 0.5mm,shorten <= 0.5mm] (6) -- (5);
  \draw[-latex,shorten >= 0.5mm,shorten <= 0.5mm] (6) -- (7);
  \draw[-latex,shorten >= 0.5mm,shorten <= 0.5mm] (7) -- (8);
  \draw[-latex,shorten >= 0.5mm,shorten <= 0.5mm,dotted] (8) -- (9);
  \draw[-latex,shorten >= 0.5mm,shorten <= 0.5mm] (9) -- (10);
\end{tikzpicture}

      \caption{Structure $\reld_{i,n}$}
      \label{fig:gadget}
    \end{figure}

The following fact is straightforward to verify.

\begin{prop}
\label{prop:unique-hom-to-N}
Let $n \geq 1$, and let $I \subseteq [n]$ be non-empty.
For each $m \in \N$, 
there is a unique homomorphism $h$ from $\reld_{I,n}$
to $\N$ such that $h(d) = m$,
namely, the homomorphism sending
each element of the form $p_i^k$ or $q_i^k$ to
$m + k$.
\end{prop}

We now establish a lemma showing that homomorphisms
between certain structures of the described form
must fix the element $d$.

\begin{lemma}
\label{lemma:homs-fix-d}
Let $n \geq 1$, let $i \in [n]$, and let $J \subseteq [n]$ be non-empty.
If there exists a homomorphism $g$ from 
$\reld_{i,n}$ to $\reld_{J,n}$, then $g(d) = d$.
\end{lemma}

\begin{proof}
Let $h$ be the unique homomorphism
(given by Proposition~\ref{prop:unique-hom-to-N})
from $\reld_{J,n}$ to $\N$ such that $h(d) = 0$;
observe that the image of $h$ is $\{ 0, \ldots, n \}$,
and that $d$ is the only element sent to $0$ by $h$.
Assume for a contradiction that $g$
is a homomorphism from $\reld_{i,n}$ to $\reld_{J,n}$
such that $g(d)$ is not equal to $d$.
Set $m = h(g(d))$; it then holds that $m > 0$.
But the composition $h(g)$ is a homomorphism
from $\reld_{i,n}$ to $\N$ sending $d$ to $m$
so by the description given by 
Proposition~\ref{prop:unique-hom-to-N},
the image of this homomorphism $h(g)$ contains $m+n$,
which is strictly greater than $n$.
This contradicts that the image of $h$ is
$\{ 0, \ldots, n \}$.
\end{proof}

Let $n \geq 1$.
For each $i \in [n]$, 
we define the formula 
$\widetilde{\theta_{i,n}}(d)$ as
$\exists p_i^1 \ldots \exists p_i^i
 \exists q_i^{i-1} \ldots \exists q_i^{n+1}
 \bigwedge_{(u,v) \in E^{\reld_{i,n}}} E(u,v)$.
We have that $\widetilde{\theta_{i,n}}(d)$ is logically equivalent to a
two-variable sentence $\theta_{i,n}(d) \in \PP{2}$ of the same size;
indeed, since the formula $\widetilde{\theta_{i,n}}(d)$
is derived from the structure $\reld_{i,n}$ which has a path-like shape,
a rewriting similar to that performed in the introduction
(Section~\ref{sect:introduction}) for the path query can be performed 
to obtain $\theta_{i,n}(d)$ from $\widetilde{\theta_{i,n}}(d)$.

For each non-empty $I \subseteq [n]$,
we define $\theta_{I,n}(d)$ to be the conjunction
$\bigwedge_{i \in I} \theta_{i,n}$.
It is well-known and straightforward to verify 
that, for any structure $\relb$,
it holds that $b \in \theta_{i,n}(\relb)$
if and only if 
there exists a homomorphism
$h$ from $\reld_{i,n}$ to $\relb$ such that $h(d) = b$.
Consequently, it holds that $b \in \theta_{I,n}(\relb)$
if and only if for every $i \in I$,
there exists a homomorphism
$h$ from $\reld_{i,n}$ to $\relb$ such that $h(d) = b$.

The following lemma is thus directly implied by
Lemma~\ref{lemma:homs-fix-d}.

\begin{lemma}
\label{lemma:theta-evaluation-d-or-empty}
Let $n \geq 1$, and let $I, J \subseteq [n]$ be non-empty.
It holds that $\theta_{I,n}(\reld_{J,n})$
is equal to $\{ d \}$ or $\emptyset$.
\end{lemma}

Whereas the just-given lemma gives us information about
the possible values of $\theta_{I,n}(\reld_{J,n})$,
the following lemma precisely characterizes
this evaluation in terms of how $I$ and $J$ are related.

\begin{lemma}
\label{lemma:theta-evaluation}
Let $n \geq 1$, and let $I, J \subseteq [n]$ be non-empty.
It holds that $\theta_{I,n}(\reld_{J,n})$
is equal to $\{ d \}$ if $I \subseteq J$,
and is equal to $\emptyset$ otherwise.
\end{lemma}

\begin{proof}
It is sufficient to prove that 
$\theta_{i,n}(\reld_{J,n})$ is equal to $\{ d \}$
if $i \in J$, and to $\emptyset$ otherwise.

In the case that $i \in J$, the identity map
is a homomorphism from $\reld_{i,n}$ to $\reld_{i,n}$,
and is hence a homomorphism from
$\reld_{i,n}$ to $\reld_{J,n}$, and so 
$d \in \theta_{i,n}(\reld_{J,n})$; 
by Lemma~\ref{lemma:theta-evaluation-d-or-empty},
it follows that
$\theta_{i,n}(\reld_{J,n}) = \{ d \}$.

In the case that $i \notin J$, we show 
that there is no homomorphism $g$ from
$\reld_{i,n}$ to $\reld_{J,n}$.
We prove the contrapositive; suppose
that there is such a homomorphism $g$.
This homomorphism has $g(d) = d$
by Lemma~\ref{lemma:homs-fix-d}.
In $\reld_{i,n}$, 
the element $d$ has an outgoing edge to $p_i^1$;
since $d$ only has outgoing edges 
in $\reld_{J,n}$ to elements of the form
$p_j^1$, with $j \in J$,
it holds that there exists $j \in J$
with $h(p_i^1) = p_j^1$.
Since in $\reld_{i,n}$,
the element $d$ has a directed path of edge length 
$i$ coming out of it, this directed path must
be mapped under $h$ to a directed path of the same length.
Observe that each vertex of the form $p_{\ell}^k$
only has outgoing edges to vertices of the same form;
so, it holds that
$h(p_i^2) = p_j^2, \ldots, h(p_i^i) = p_j^i$
and we have that $i \leq j$.
But if $q_i^{i-1}$ is mapped under $h$ to 
$p_j^{i-1}$, then by the observation, 
$q_i^i$ is mapped under $h$ to $p_j^i$;
since $q_i^i$ has a directed path of edge length
$n+1-i$ coming out of it, its image $p_j^i$ must as well;
but this is a contradiction, since by the definition
of the $\reld_{j,n}$ no point $p_j^i$ has
such a directed path (this would imply a directed
path of edge length $n+1$ coming out of $d$ in $\reld_{j,n}$).
So $q_i^{i-1}$ must be mapped under $h$ to 
a point in $\reld_{j,n}$ not equal to $p_j^{i-1}$
but having an edge to $h(p_i^i) = p_j^i$;
this point must be $q_{j}^{j-1}$, and so we conclude
that $i = j$.
\end{proof}

The following theorem shows how
one can simulate unary symbols with a binary symbol.

\begin{theorem} 
\label{thm:simulating-unary-by-binary}
There exists a polynomial $p$ such that,
for any $m, s \geq 1$,
there exist 
\begin{itemize}
\item  a set $B^+ \supseteq [s]$ of size 
less than or equal to $p(m,s)$,
\item  formulas $\psi_1(d), \ldots, \psi_m(d)\in\EP{2}_{\{E\}}$,
and
\item  a mapping $t$ from the set 
$$\{ E(v,w) ~|~ v, w \in B^+ \}$$ 
to the set
$$\{ U_i(j) ~|~ i \in [m], j \in [s] \} \cup \{ \bot, \top \}$$
\end{itemize}
such that the following holds.
Let $\relb$ be any $\sigma$-structure with universe $[s]$ and
$\sigma \supseteq \{U_i\mid i\in[m]\}$;
when $\relb'$ is defined as the $\{E\}$-structure
with universe $B^+$
and where 
$$\relb' \models E(v,w) \textup{ if and only if }
\relb \models t(E(v,w)),$$
it holds that (for each $i \in [m]$)
$\psi_i(\relb') = U_i^{\relb}$. 
\end{theorem}

\begin{proof}
We design the assignment $t$ such that,
for any structure $\relb$ of the described form,
the structure $\relb'$ will be as follows.
The universe $B^+$ of $\relb'$
is $\cup_{i \in [m]} D_{i,m+1} \times [s]$.
However, for each $j \in [s]$
we refer to the element $(d,j)$ using the name $j$ as well,
so in this way we have $[s] \subseteq B^+$.
\begin{itemize}
\item 
For each $j \in [s]$, define $I_j$ to be
$\{ i \in [m] ~|~ j \in U_i^{\relb} \} \cup \{ m+1 \}$,
that is, $I_j$ contains $m+1$ and
the subscripts of all unary symbols
$U_i$ such that $\relb \models U_i(j)$.

\item  For $I \subseteq [m+1]$ and $j \in [s]$, define
$\reld_{I,m+1} \times \{ j \}$ as the structure
obtained from $\reld_{I,m+1}$ by renaming each element
$e$ as $(e,j)$.

\item  Define $E^{\relb'}$
as the union
$\bigcup_{j \in [s]} E^{\reld_{I_j,m+1} \times \{ j \}}$.

\end{itemize}
From this description, it is evident that 
the desired assignment $t$ exists:
for $j \in [s]$,
the edges in $E^{\reld_{m+1,m+1} \times \{ j \}}$
are always included in $E^{\relb'}$;
for $j \in [s]$ and
$i \in [m]$, the edges in
$E^{\reld_{i,m+1} \times \{ j \}}$
are included in $E^{\relb'}$ depending on whether or not
$\relb \models U_i(j)$,
and all other potential edges are never included
in $E^{\relb'}$.

For each $i \in [m]$, define 
$\psi_i(d) = \theta_{i,m+1}(d)$.
To conclude the proof, we thus verify that
$\psi_i(\relb') = U_i^{\relb}$.
We have
\begin{align}
\psi_i(\relb') 
&= \cup_{j \in [s]} \psi_i(\reld_{I_j,m+1} \times \{ j \}) \\
&= \cup_{j \in [s]} \{ (d,j) ~|~ 
    d \in \psi_i(\reld_{I_j,m+1}) \} \\
&= \cup_{j \in [s]} \{ (d,j) ~|~ 
    d \in \theta_{i,m+1}(\reld_{I_j,m+1}) \} \\
&= \cup_{j \in [s]} \{ (d,j) ~|~
    i \in I_j \} \mbox{\hspace{4pt} (by Lemma~\ref{lemma:theta-evaluation})} \\
&= \cup_{j \in [s]} \{ j ~|~ i \in I_j \}
 = U_i^{\relb}.
\end{align}
\end{proof}

Now we are ready to prove the compilation lower bound for signatures
that contain at least one at least binary symbol.

\begin{proof}[Proof of Theorem~\ref{thm:lower-bound-at-least-binary}]
  Since an $r$-ary relation can be used to simulate a binary relation, it suffices to prove the theorem for the signature $\sigma=\{E\}$
  that contains a single binary symbol $E$.

  The proof follows closely and makes use of the proof of
  Theorem~\ref{thm:lower-bound-inf-prop-two-unary}; 
  in particular, we make use of the case of this theorem
  when the signature contains infinitely many unary symbols.
  We will use
  Theorem~\ref{thm:simulating-unary-by-binary} to simulate unary
  relations by a single binary relation.

  Consider the sentence $\varphi_{n,k}$ from the proof of
  Theorem~\ref{thm:lower-bound-inf-prop-two-unary}
  (adjusted accordingly so that all symbols are unary). 
  By renaming the unary
  symbols we can assume that $\varphi_{n,k}$ is a sentence over the
  signature 
  $$\sigunary=\{U_i\mid i\in[m]\}$$ 
  where $m=n^2+n$.
  By applying Theorem~\ref{thm:simulating-unary-by-binary} for
  $m=n^2+n$ and $s=n$, we obtain formulas 
  $$\psi_1(d), \ldots, \psi_m(d)\in\EP{2}_{\sigma}$$ 
  that can be used to simulate the unary
  symbol in $\sigunary$.
  Let $\varphi^{\sigma}_{n,k}$ be the existential-positive
  $\sigma$-sentence obtained by replacing every occurrence of an
  atom $U_i(x)$ in $\varphi_{n,k}$ by $\psi_i(x)$
  (from Theorem~\ref{thm:simulating-unary-by-binary}).
  Note that $\varphi^{\sigma}_{n,k}\in \exEP{2}_\sigma$ because
  (by Proposition~\ref{prop:unarycompilestoEPone}) there is a sentence
  $\psi_{n,k}\in \EP{1}_\sigunary$ that is logically equivalent
  to $\varphi_{n,k}$ and which can be transformed to an $\EP{2}_\sigma$-sentence that is logically equivalent to $\varphi^{\sigma}_{n,k}$ by
  the same replacement of atoms $U_i(x)$ by $\psi_i(x)$.

  To prove the lower bound let $\psi\in \PFO{\ell}_\sigma$
  be a sentence that is logically equivalent to
  $\varphi^{\sigma}_{n,k}$ for $k=\bigl\lfloor \tfrac14(n/\log
  n)^{2/3}\bigr\rfloor$. Let $B^+$ be the domain and $t$ the mapping
  provided by Theorem~\ref{thm:simulating-unary-by-binary}.
  By
  Lemma~\ref{lem:formulas-to-circuits} and the polynomial bound on $|B^+|$ there exists a monotone
  circuit $C^\sigma_{|B^+|}$ of size polynomial in $n$ and $\|\psi\|$ that
  evaluates $\psi$ on $\sigma$-structures over the universe $B^+$. Now
  we connect the inputs of this circuit to the inputs for
  $\sigunary$-structures over the universe $[n]$ and $\bot, \top$ according to the
  mapping $t$.
  From Theorem~\ref{thm:simulating-unary-by-binary} and the definition
  of $\varphi^{\sigma}_{n,k}$ it follows that
  the resulting circuit $C^\sigunary_{n}$, which is of size polynomial
  in $n$ and $\|\psi\|$, decides whether a $\sigunary$-structure $\relb$ with
  universe $[n]$ satisfies $\varphi_{n,k}$. 
  As in the proof of Theorem~\ref{thm:lower-bound-inf-prop-two-unary}, we
  can hard-code some inputs to obtain a circuit $C^{\{E\}}_{n}$ of
  the same size that decides the existence of a $k$-clique on
  $n$-vertex graphs. By applying Theorem~\ref{thm:monotone-clique} we
  obtain a lower bound of  $2^{\Omega((n/\log n)^{1/3})}$ on the size of this
  circuit which is on the other hand bounded by a polynomial in
  $\|\phi\|$ and $n$. Since $N=\|\varphi^{\sigma}_{n,k}\|$ is
  polynomial in $n$ it follows that $\|\psi\|=\Omega(2^{N^\delta})$ for some $\delta > 0$
  and the theorem is proven.
\end{proof}

\begin{theorem}
\label{thm:chopped-np-at-least-bin}
When $\sigma$ is an  signature that contains an at least binary symbol,
the parameterized problem 
$$\qe(\exEP{2}_\sigma)$$
is $\chopped{\np}$-hard.
\end{theorem}

\begin{proof}
The proof is similar to the proof of Theorem~\ref{thm:chopped-np-two-distinct-non-prop}.
In that proof, we used the fact that
$(\graphG,k)$ is a yes instance of $\clique$
if and only if 
$\rela_\graphG \models \phi_{n,k}$;
this in turn holds if and only if
$\rela'_\graphG \models \phi_{n,k}^\sigma$,
where $\rela'_\graphG$ is obtained from $\rela_\graphG$
as described in the statement of Theorem~\ref{thm:simulating-unary-by-binary}.
As in the proof of Theorem~\ref{thm:chopped-np-two-distinct-non-prop},
it suffices to give a poly-comp reduction from $\nuclique$ to $\qe(\exEP{2}_\sigma)$.
Define $(g,s)$ as follows:
$g$ maps an instance $(\graphG,k)$ of $\clique$
to $(\phi_{n,k}^\sigma,\rela'_\graphG)$,
and $s(n) = \{ \phi_{n,i}^\sigma ~|~ i \leq n \}$.
We have that $g$ is polynomial-time computable,
and $(g,s)$ yields the desired poly-comp reduction.
\end{proof}

\section{Compilability to arbitrary formats}
\label{sect:param-comp-summary}

In the classification theorems,
for all fragments where we obtained an exponential lower bound
on the size of compilation to $\PFO{\ell}$,
we also can conditionally rule out compilations to arbitrary polynomial-size formats.
The results can be formulated as follows.

\begin{theorem}
\label{thm:non-param-comp}
Assume that the polynomial hierarchy does not collapse; 
then, the following hold.
\begin{itemize}
  \item 
When $\sigma$ is an infinite signature that contains two distinct non-propositional symbols, it holds that
the problem $\qe(\exEP{1}_\sigma)$ is not in $\polycompp$.

\item When $\sigma$ is a signature that contains an at least binary symbol,
it holds that 
the problem $\qe(\exEP{2}_\sigma)$ is not in $\polycompp$.
\end{itemize}
\end{theorem}

\begin{proof}
The statements follow from 
Theorems~\ref{thm:chopped-np-two-distinct-non-prop}
and~\ref{thm:chopped-np-at-least-bin}, in conjunction with
Theorem~\ref{thm:chopped-np-not-in-polycompp}.
\end{proof}

Results of this form were previously shown by the second author~\cite[Theorem 4.3]{Chen-tocl}.
However, the present theorem, as indicated, is more comprehensive in that it provides
a tight analysis: 
it provides a negative result for each value $k$ and each signature $\sigma$
such that $\exEP{k}_\sigma$ did not have a polynomial-size compilation
according to our classification theorems.  Also, the present article
permits propositional symbols, whereas in the article~\cite{Chen-tocl},
all symbols were assumed to be of arity $1$ or greater.

Let us offer some discussion concerning the interpretation of
Theorem~\ref{thm:non-param-comp}.
A parameterized problem $(Q,\kappa)$ is in $\polycompp$
if there exists a computable, polynomial length function $c$
such that when a string $x$ is presented along with $c(\kappa(x))$,
whether or not $x$ is in $Q$ can be decided in polynomial time.\footnote{
  We remark that, for the poly-compilable function $g$ 
  in the definition of \emph{poly-compilable to $\fancyc$},
  it is straightforwardly verifiable
  that one can always take $f$ to be the identity map; the described
  characterization follows from this remark.
}
In the case of one of our problems $\qe(\fancyl)$, since
the parameter of an instance $(\phi,\relb)$ is the sentence $\phi$,
this means that one should consider the situation where each instance
$(\phi,\relb)$
is presented along with $c(\phi)$.
The theorem given above rules out (for the fragments presented) the possibility of a 
computable, polynomial length function $c$ with the described properties.
That is, it in essence rules out polynomial length compilations
to an arbitrary format that consequently allow for polynomial-time decision.
This ruling out, however, is under the complexity-theoretic assumption
that the polynomial hierarchy does not collapse.
It can be contrasted with our unconditional results: for those,
no complexity-theoretic assumption is invoked, but what is shown is that
compiling to a \emph{specific} format, namely that of bounded-variable positive
first-order queries, requires exponential size.

{\bf Acknowledgements.}
The collaboration leading to this paper was partly carried out during the
 special semester program on Computational and Proof Complexity 
 held in St. Petersburg in 2016.
 The authors are grateful for having had the opportunity to attend,
 and thank the organizers 
Edward A. Hirsch and Sam Buss, as well as the rest of the program organization.

\bibliographystyle{ACM-Reference-Format}
\balance
\bibliography{bibliography}

\end{document}